# Dynamic State Estimation of Power System Utilizing Cauchy Kernel-Based Maximum Mixture Correntropy UKF over Beluga Whale-Bat Optimization

Duc Viet Nguyen, Haiquan Zhao, *Senior Member, IEEE*, Jinhui Hu

*Abstract*— Non-Gaussian noise, outliers, sudden load changes, and bad measurement data are key factors that diminish the accuracy of dynamic state estimation in power systems. Additionally, unscented Kalman filters (UKF) based on correntropy criteria utilize bandwidth-sensitive Gaussian kernels, which may lead to singular matrices in the Cholesky decomposition. To overcome all the above problems, in this paper, a robust UKF based on Cauchy kernel maximum mixture correntropy (CKMMC) criteria over hybrid Beluga Whale-Bat (BWB) optimization (BWB-CKMMC-UKF) is proposed, in which the kernel is merged of two Cauchy functions. Specifically, the measurement error and state error are unified in the cost function by the statistical linearization technique, and the optimal value of state estimation is obtained by fixed-point iteration. Because of its insensitive feature to kernel bandwidth and notable thick-tailed feature, the Cauchy kernel function is utilized instead of the Gaussian kernel in the optimization criteria. Additionally, to fit the power system model, the shape coefficients of the kernel in the CKMMC criterion and scale coefficients that influence the selection of sigma points in the unscented transform are determined based on the BWB algorithm. Simulation results on IEEE 14, 30, and 57-bus test systems validated the performance of the proposed algorithm.

## I. INTRODUCTION

THE accurate dynamic state estimation (DSE) method is crucial in the monitoring, management, and control of power systems. In practice, power systems are always affected by complex factors such as random outliers, instrument failures, sudden load changes, etc [1]. Currently, state estimation methods can be grouped into two main categories: model-based and data-driven approaches, which are the most prominent techniques being widely applied [2]. Model-based estimation methods, such as the Kalman filter, are recognized as recursive algorithms that offer real-time processing capabilities and relatively low computational complexity [3]. However, these methods rely heavily on the accuracy of the dynamic model and prior system knowledge to adequately define the associated parameters. On the other hand, data-driven estimation methods, such as deep neural networks, which often employ convolutional neural networks, have demonstrated competitive accuracy and robustness compared to model-based approaches [4,5]. Nevertheless, they depend on the quality and coverage of the training dataset, as well as memory requirements [6]. As a result, neural network-based methods often lack robustness and interpretability when faced with unseen operating conditions [7]. A current design trend is the integration of both aforementioned approaches, which aims to enhance interpretability while maintaining data efficiency and implicitly learning complex dynamics from the available data [2,6,7].

In this paper, our objective is to focus on the analysis and evaluation of current limitations inherent in model-based methods-specifically, the Kalman filter-and accordingly propose solutions to overcome their restrictive constraints. The unscented Kalman filter (UKF), known for its computational efficiency, has been widely employed in dynamic state estimation of power systems (DSE-PS). However, the UKF's reliance on the unscented transform with Gaussian assumptions may not accurately represent modern power systems' increasingly complex and unpredictable nature [8,9]. Similarly, the extended Kalman filter (EKF), which utilizes first-order Taylor expansion for estimation accuracy, often falls short of delivering optimal performance. Surveying existing research, several prominent algorithms have been introduced to tackle these difficult problems. To address the problem of measurement noise with unknown statistics, a cubature Kalman filtering scheme is proposed in [10], which dynamically adjusts its parameters based on the reduced Gaussian mixture model obtained from the sliding window. A novel estimator is designed in [11] based on the encoding-decoding mechanism. However, these algorithms still have some limitations, such as dependence on the quality of measurement data, the value of free parameters, or the ability to handle uncertain factors on objects that have not been verified.

The advent of information-theoretic learning has led to incorporating correntropy criteria in Kalman filtering, addressing limitations inherent in the minimum mean square error (MMSE) approach [12,13]. This innovation has resulted in enhanced estimation performance for nonlinear systems, particularly through the maximum correntropy criterion Kalman filter (MCC-KF) [14,15]. Nevertheless, the fixed Gaussian kernel shape in MCC criteria has shown limitations in diverse scenarios [16]. Recent research has introduced a robust variant of MCC-UKF, incorporating methods to

This work was supported by the National Natural Science Foundation of China (grant: 62171388, 61871461, 61571374). (*Corresponding author*: Haiquan Zhao).

Duc Viet Nguyen, Haiquan Zhao, and Jinhui Hu are with the Key Laboratory of Magnetic Suspension Technology and Maglev Vehicle, Ministry of Education, School of Electrical Engineering Southwest Jiaotong University Chengdu, China. (e-mail: ruandeyue@my.swjtu.edu.cn; hqzhao_swjtu@126.com; jhhu_swjtu@126.com).

enhance numerical stability [17]. This development represents a significant step towards improving the adaptability and reliability of state estimation techniques in complex power systems.

It is easy to observe that correntropy-based filters predominantly employ Gaussian kernels to quantify vector distances. However, this approach presents limitations when dealing with multidimensional, non-Gaussian noise in practical applications. The Gaussian kernel-based Cholesky decomposition often encounters singular matrices, leading to computational instability [18]. Moreover, determining the optimal kernel bandwidth for the MCC-UKF typically requires extensive experimentation and fine-tuning [19]. To address these constraints, researchers have introduced Cauchy kernel-based maximum correntropy (CKMC) [20,21,22]. The Cauchy kernel demonstrates superior robustness compared to its Gaussian counterpart, particularly when handling data with non-uniform distributions [22]. Additionally, the concept of mixture correntropy, which combines various Gaussian functions, has been proposed to mitigate issues related to affected data attributes [23,24]. Building upon these advancements, a study on the fusion of two Cauchy kernels is proposed, termed Cauchy kernel-based maximum mixture correntropy (CKMMC). This approach aims to leverage the strengths of both the Cauchy kernel and mixture correntropy to enhance the performance and robustness of correntropy-based filtering techniques.

However, the incorporation of dual Cauchy kernels introduces additional complexity through an increased number of kernel shape coefficients. Furthermore, the selection of scale coefficients, which influence sigma point determination in the unscented transform (UT), requires careful consideration [25,26]. These coefficients play a critical role in algorithm performance, and their optimal selection remains a significant challenge [24,25,26]. The estimation process frequently relies on extensive experimentation and the researcher's expertise concerning the dynamics of the system under estimation. Recent literature has extensively examined the impact of these coefficients on estimation algorithm performance. Based on this critical issue, some studies have introduced kernels with adaptive bandwidths [27]. However, these methods only solve part of the problem and do not provide a comprehensive solution, since scale coefficients also affect the choice of sigma point in the unscented transform. Meta-heuristic optimization techniques have emerged as a preferred approach for tackling the challenge of coefficient optimization and identifying missing parameters in object models [26,28,29]. These techniques, which offer significant potential for improving the robustness and efficiency of estimation algorithms in complex systems, are particularly valuable where traditional methods fall short. However, the original optimization algorithms often suffer from some limitations related to search ability, convergence speed, or failure to achieve global optimization, which can be solved through hybridization or the application of mathematical functions. Therefore, constructing a hybrid meta-heuristic algorithm, which combines the advantages of each method, is a simple yet effective way to simultaneously address both the challenges of determining coefficient values and overcome the limitations of single optimization algorithms.

In summary, non-Gaussian noise and outliers, sudden load change, bad measurement data, the appearance of singular matrices in the Cholesky decomposition of the Gaussian kernel, and the optimal value of coefficients are the main motivations in this paper. Therefore, a robust estimation algorithm, termed BWB-CKMMC-UKF, is proposed to overcome all the above limitations. The main contributions are as:

1) Two Cauchy kernel functions with a scaling coefficient are merged to obtain the mixture correntropy as the optimization criteria (CKMMC), which increases flexibility, is insensitive to kernel bandwidth, and solves the problem of singular matrices.

2) A UKF based on CKMMC criteria (CKMMC-UKF) for DSE-PS is developed.

3) A hybrid meta-heuristic Begula Whale-Bat (BWB) optimization algorithm is designed. The advantages of the Beluga Whale optimization and the Bat algorithms combined enhance the accuracy and convergence speed of the optimization process. BWB can automatically select the optimal value of coefficients and matrices or find the missing parameters of the object model.

4) A robust BWB-CKMMC-UKF algorithm for DSE-PS is proposed that can overcome all the limitations analyzed above. In which to fit with the power systems model, the optimal value of the coefficients, such as the shape coefficients of the kernel in CKMMC criteria and the scale coefficients that impact the selection of sigma points in unscented transform are determined based on the BWB algorithm, thereby enhancing the estimation performance.

This paper consists of six main parts: Part II describes the PS model and CKMMC criteria; Part III derives the CKMMC-UKF algorithm; Part IV derives the BWB-CKMMC-UKF; Part V reports experimental results; and Part VI concludes.

## II. POWER SYSTEM MODEL AND CAUCHY KERNEL-BASED MAXIMUM MIXTURE CORRENTROPY CRITERIA

### A. Power System Dynamic Model

In order to DSE-PS, a physical and mathematical model are established [1,3,24,30], which consists of dynamic nonlinear equations between measurements and state variables described as follows:

$$\begin{cases} \mathbf{x}_t = \mathbf{f}\left(\mathbf{x}_{t-1}\right) + \mathbf{q}_t \\ \mathbf{y}_t = \mathbf{g}\left(\mathbf{x}_t\right) + \mathbf{r}_t \end{cases} \tag{1}$$

where $\mathbf{x}_t$: the state variable vector dimension $n$ and contains the angle and magnitudes of each node at time $t$; $\mathbf{y}_t$: the measurement vector dimension $m$ at time $t$ and contains voltage magnitude measurements, real power flow measurements, real power injection measurements, reactive power flow measurements, and reactive power injection measurements; $\mathbf{f}(\mathbf{x}_{t-1})$: state transition function of $\mathbf{x}_{t-1}$; $\mathbf{g}(\mathbf{x}_t)$:

measurement function of $\mathbf{x}_t$; $\mathbf{r}_t$ and $\mathbf{q}_t$ represent measurement noise, process noise at time $t$ with covariance matrices $\mathbf{R}_t \in \mathbb{R}^{m \times m}$ and $\mathbf{Q}_t \in \mathbb{R}^{n \times n}$, respectively.

The state transition function $\mathbf{f}(\mathbf{x}_{t-1})$ can be represented using a state prediction method. By employing Holt's two-parameter exponential smoothing technique [3,24], the function $\mathbf{f}(\mathbf{x}_{t-1})$ is derived as follows:

$$\mathbf{f}\left(\mathbf{x}_{t-1}\right) = \mathbf{\Delta}_{t-1} + \mathbf{\Gamma}_{t-1} \tag{2}$$

$$\mathbf{\Delta}_{t-1} = \Upsilon \mathbf{x}_{t-1} + \left(1 - \Upsilon\right) \tilde{\mathbf{x}}_{t-1} \tag{3}$$

$$\mathbf{\Gamma}_{t-1} = \Upsilon \left(\mathbf{\Delta}_{t-1} - \mathbf{\Delta}_{t-2}\right) + \left(1 - \theta\right) \mathbf{\Gamma}_{t-2} \tag{4}$$

where $\Upsilon \in (0,1)$ ; $\theta \in (0,1)$ ; $\mathbf{x}_{t-1}$ , $\tilde{\mathbf{x}}_{t-1}$ : state vectors, predicted state vector respectively at time $t-1$.

In addition, the measurement function $\mathbf{g}(\mathbf{x}_t)$ represents the real power relationship at time $t$: the standard real, line flow, and reactive power balance equations, which are described by the following equations:

$$\begin{cases} P_{i,t} = \sum_{j=1}^{N} |V_{i,t}||V_{j,t}| \left(S_{ij} \cos \varphi_{ij,t} + F_{ij} \sin \varphi_{ij,t}\right) \\ Q_{i,t} = \sum_{j=1}^{N} |V_{i,t}||V_{j,t}| \left(S_{ij} \sin \varphi_{ij,t} - F_{ij} \cos \varphi_{ij,t}\right) \\ P_{ij,t} = V_{i,t}^2 \left(S_{gi} + S_{ij}\right) - |V_{i,t}||V_{j,t}| \left(S_{ij} \cos \varphi_{ij,t} + F_{ij} \sin \varphi_{ij,t}\right) \\ Q_{ij,t} = -V_{i,t}^2 \left(F_{gi} + F_{ij}\right) - |V_{i,t}||V_{j,t}| \left(S_{ij} \sin \varphi_{ij,t} - F_{ij} \cos \varphi_{ij,t}\right) \end{cases} \tag{5}$$

where $\varphi_{ij,t} = \varphi_{i,t} - \varphi_{j,t}$: the voltage phase difference at time $t$ between the $i$ and $j$ nodes; $F_{gi}$ and $S_{gi}$ represent the susceptance and conductance of the Shunt at bus $i$; $F_{ij}$ and $S_{ij}$ represent the susceptance and conductance of the line between the $i$ and $j$ nodes; $Q_{i,t}$ and $P_{i,t}$: represent the reactive power injection and real power injection at time $t$ of node $i$; $Q_{ij,t}$ and $P_{ij,t}$: represent the reactive power flow and real power flow at time $t$ between nodes $i$ and $j$, respectively; $|V|$ denote the voltage magnitude.

### B. CKMMC criteria

Considering two random variables $\mathbf{X}$ and $\mathbf{Y}$, and according to [12,13,14,15,18,19] the correntropy is defined as follows:

$$V(\mathbf{X}, \mathbf{Y}) = \mathbf{E}\left[\kappa(\mathbf{X}, \mathbf{Y})\right] = \int \kappa(x, y) d\mathbf{F}_{X,Y}(x, y) \tag{6}$$

where $\mathbf{E}$: the expectation operation; $\kappa$ : Mercer kernel function; $\mathbf{F}_{X,Y}(x, y)$ : the joint distribution function of $\mathbf{X}$ and $\mathbf{Y}$.

According to [23,24], the mixture correntropy is defined as:

$$M(\mathbf{X}, \mathbf{Y}) = \mathbf{E}\left[\delta_1 \kappa_1(\mathbf{X}, \mathbf{Y}) + \delta_2 \kappa_2(\mathbf{X}, \mathbf{Y})\right] \tag{7}$$

where $\delta$: mixture coefficient ($0 < \delta < 1$ and $\delta_1 + \delta_2 = 1$); $\kappa_1$, $\kappa_2$: denotes two different kernel. According to the sample mean estimator, Eq.(7) can be calculated as follows:

$$M(\mathbf{X}, \mathbf{Y}) = \frac{1}{N} \sum_{i=1}^{N} \left[\delta_1 \kappa_1(x_i, y_i) + \delta_2 \kappa_2(x_i, y_i)\right] \tag{8}$$

where: $x_i$, $y_i$: the $i^{th}$ element of the random variables $\mathbf{X}$ and $\mathbf{Y}$, respectively; $N$: the number of samples. With the advantages that have been analyzed, this paper will use the Cauchy kernel function when constructing the optimality criteria. Consider the problem of using two or more kernels, which brings both advantages and challenges. Increasing the number of kernels will lead to an increase in computational complexity and convergence of the algorithm. Therefore, for simplicity, this paper examines the case of combining two Cauchy kernels. The Cauchy kernel function is given by [18,19,20,21,22]:

$$C_\sigma(e) = \frac{1}{1 + e^2 / \sigma} \tag{9}$$

where $C_\sigma$ : Cauchy kernel function; $e = x - y$ ; $\sigma$: kernel shape coefficients ($\sigma > 0$). Combining Eq.(9) and (8), Eq. (8) can be written as:

$$M(\mathbf{X}, \mathbf{Y}) = \frac{1}{N} \sum_{i=1}^{N} \left[\delta_1 \frac{1}{1 + e_i^2 / \sigma_1} + \delta_2 \frac{1}{1 + e_i^2 / \sigma_2}\right] \tag{10}$$

It can be seen that Eq.(10) achieves its maximum value when the error $e_i$ is 0. This is the CKMMC criteria mentioned in this paper.

## III. DERIVATION OF THE CKMMC-UKF ALGORITHM

This section mainly focuses on researching CKMMC-UKF and giving the specific step-by-step formula derivation.

### A. Time update

Derived from state estimation $\mathbf{x}_{t-1|t-1}$ , estimate error covariance matrix $\mathbf{P}_{t-1|t-1}$ and according to the UT transform, obtained $2n+1$ sigma points at time $t-1$ as follows [3,8,17]:

$$\psi_{t-1|t-1}^{\ell} = \begin{cases} \hat{\mathbf{x}}_{t-1|t-1} & ; \ell = 0 \\ \hat{\mathbf{x}}_{t-1|t-1} + \left(\sqrt{(n+\zeta)\mathbf{P}_{t-1|t-1}}\right)_\ell & ; \ell = 1, 2, .., n \\ \hat{\mathbf{x}}_{t-1|t-1} - \left(\sqrt{(n+\zeta)\mathbf{P}_{t-1|t-1}}\right)_{\ell-n} & ; \ell = n+1, ..., 2n \end{cases} \tag{11}$$

$$\zeta = \omega^2 \left(n + \lambda\right) - n \tag{12}$$

where $\omega$: the distribution coefficient of sigma points; $\left(\sqrt{(n+\zeta)\mathbf{P}_{t-1|t-1}}\right)$ denotes the $\ell^{th}$ column of matrix $\sqrt{(n+\zeta)\mathbf{P}_{t-1|t-1}}$ ; $\lambda$: the scale coefficient. Normally, $\lambda = 3 - n$ when the state variable is multivariable and set $\lambda = 0$ in the non-Gaussian environment [24].

Then, $\hat{\mathbf{x}}_{t|t-1}$ and $\mathbf{P}_{t|t-1}$ represents the prior state estimation and the prior state error covariance matrix are computed:

$$\hat{\mathbf{x}}_{t|t-1} = \sum_{\ell=0}^{2n} \phi_b^\ell \mathbf{f}\left(\psi_{t-1,t-1}^\ell\right) \tag{13}$$

$$\mathbf{P}_{t|t-1} = \sum_{\ell=0}^{2n} \phi_c^\ell \left(\mathbf{f}\left(\psi_{t-1,t-1}^\ell\right) - \hat{\mathbf{x}}_{t|t-1}\right)\left(\mathbf{f}\left(\psi_{t-1,t-1}^\ell\right) - \hat{\mathbf{x}}_{t|t-1}\right)^T + \mathbf{Q}_{t-1} \tag{14}$$

where: $\phi_c^\ell$ : the variance weight, $\phi_b^\ell$ : mean weight and can be described as follows:

$$\phi_b^0 = \frac{\zeta}{n + \zeta} \tag{15}$$

$$\phi_c^0 = \frac{\zeta}{n + \zeta} + \left(1 - \alpha^2 + \beta\right) \tag{16}$$

$$\phi_b^\ell = \phi_c^\ell = \frac{1}{2\left(n + \zeta\right)}; \quad \ell = 1, 2, ..., 2n \tag{17}$$

where: $\alpha$, $\beta$: the factors that affect the process of obtaining $2n+1$ sigma points.

*B. Measurement Update:* Continue to perform the unscented transform on $\hat{\mathbf{x}}_{t|t-1}$ and $\mathbf{P}_{t|t-1}$ to obtain $2n+1$ sigma point:

$$\psi_{t|t-1}^{\ell} = \begin{cases} \hat{\mathbf{x}}_{t|t-1} & ; \ell = 0 \\ \hat{\mathbf{x}}_{t|t-1} + \left(\sqrt{(n+\zeta)\mathbf{P}_{t|t-1}}\right)_{\ell} & ; \ell = 1,2,...,n \\ \hat{\mathbf{x}}_{t|t-1} - \left(\sqrt{(n+\zeta)\mathbf{P}_{t|t-1}}\right)_{\ell-n} & ; \ell = n+1,...,2n \end{cases} \quad (18)$$

Then, $\mathbf{P}_{xy,t}$ and $\hat{\mathbf{y}}_{t|t-1}$ the cross-covariance matrix and prior mean of measurement, respectively, are calculated by the following formulas:

$$\hat{\mathbf{y}}_{t|t-1} = \sum_{\ell=0}^{2n} \phi_i^{\ell} \mathbf{g}_i(\psi_{t|t-1}^{\ell}) \quad (19)$$

$$\mathbf{P}_{xy,t} = \sum_{\ell=0}^{2n} \phi_c^{\ell} \left(\psi_{t|t-1}^{\ell} - \hat{\mathbf{x}}_{t|t-1}\right)\left(\mathbf{g}_i(\psi_{t|t-1}^{\ell}) - \hat{\mathbf{y}}_{t|t-1}\right)^T \quad (20)$$

In order to complete the measurement update process, a dual noise model consisting of measurement variables and state variables is constructed. A measurement slope matrix can be defined as:

$$\tilde{\mathbf{U}}_t = \left(\mathbf{P}_{t|t-1}^{-1}\mathbf{P}_{xy,t}\right)^T \quad (21)$$

Then, measurement (1) can be computed as follows:

$$\mathbf{y}_t \approx \hat{\mathbf{y}}_{t|t-1} + \tilde{\mathbf{U}}_t\left(\mathbf{x}_t - \hat{\mathbf{x}}_{t|t-1}\right) + \mathbf{r}_t \quad (22)$$

Here, the prior state estimation error is described as:

$$\wp_t = \mathbf{x}_t - \hat{\mathbf{x}}_{t|t-1} \quad (23)$$

Combining Eq.(22) and Eq.(23) yields:

$$\begin{bmatrix} \hat{\mathbf{x}}_{t|t-1} \\ \mathbf{y}_t - \hat{\mathbf{y}}_{t|t-1} + \tilde{\mathbf{U}}_t\hat{\mathbf{x}}_{t|t-1} \end{bmatrix} = \begin{bmatrix} \mathbf{I} \\ \tilde{\mathbf{U}}_t \end{bmatrix}\mathbf{x}_t + \chi_t \quad (24)$$

where $\mathbf{I}$: unity matrix; $\chi_t = \begin{bmatrix} -\wp_t \\ \mathbf{r}_t \end{bmatrix}$ and

$$E\left[\chi_t\chi_t^T\right] = \begin{bmatrix} \mathbf{P}_{t|t-1} & 0 \\ 0 & \mathbf{R}_t \end{bmatrix} = \begin{bmatrix} \mathbf{B}_{P,t|t-1}\mathbf{B}_{P,t|t-1}^T & 0 \\ 0 & \mathbf{B}_{R,t}\mathbf{B}_{R,t}^T \end{bmatrix} = \mathbf{B}_t\mathbf{B}_t^T \quad (25)$$

where $\mathbf{B}_{P,t|t-1} \in \mathbb{R}^{n\times n}$ and $\mathbf{B}_{R,t} \in \mathbb{R}^{m\times m}$: the Cholesky decomposition factors of $\mathbf{P}_{t|t-1}$ and $\mathbf{R}_t$, respectively. Similar to the work [3,8,13,17], the positive definite condition was assumed to be satisfied when performing the Cholesky decomposition. Continue multiplying both sides of Eq.(24) by $\mathbf{B}_t^{-1}$, obtaining:

$$\mathbf{L}_t = \mathbf{D}_t\mathbf{x}_t + \mathbf{e}_t \quad (26)$$

with $\mathbf{L}_t = \mathbf{B}_t^{-1}\begin{bmatrix} \hat{\mathbf{x}}_{t|t-1} \\ \mathbf{y}_t - \hat{\mathbf{y}}_{t|t-1} + \tilde{\mathbf{U}}_t\hat{\mathbf{x}}_{t|t-1} \end{bmatrix} = [l_{1,t}, l_{2,t},...,l_{n+m,t}]^T \quad (27)$

$$\mathbf{D}_t = \mathbf{B}_t^{-1}\begin{bmatrix} \mathbf{I} \\ \tilde{\mathbf{U}}_t \end{bmatrix} = \left[\mathbf{d}_{1,t}^T,...,\mathbf{d}_{n+m,t}^T\right]^T \quad (28)$$

$$\mathbf{e}_t = \mathbf{B}_t^{-1}\begin{bmatrix} -\wp_t \\ \mathbf{r}_t \end{bmatrix} = \left[e_{1,t},...,e_{n+m,t}\right]^T \quad (29)$$

Based on [23,24] and combining Eq.(10) and Eq.(29), the cost function of CKMMC is described as follows:

$$\mathbf{J}_{CKMMC}\left(\mathbf{x}_t\right) = \frac{1}{N}\sum_{j=1}^{2}\sum_{i=1}^{N}\left[\delta_j \frac{1}{1+e_{i,t}^2/\sigma_j}\right] \quad (30)$$

where: $e_{i,t} = l_{i,t} - \mathbf{d}_{i,t}\mathbf{x}_t$ and $N = n + m$. To find the extreme value of $J_{CKMMC}(\mathbf{x}_t)$, the first need to compute the derivative of $J_{CKMMC}(\mathbf{x}_t)$ with $\mathbf{x}_t$. The optimal estimate is the maximum value:

$$\hat{\mathbf{x}}_{t|t} = \arg\max_{\mathbf{x}_t}\left(\frac{1}{N}\sum_{j=1}^{2}\sum_{i=1}^{N}\left[\delta_j \frac{1}{1+e_{i,t}^2/\sigma_j}\right]\right) \quad (31)$$

Setting the gradient of Eq.(31) with $\mathbf{x}_t$ to be zero yields:

$$\frac{\partial J_{CKMMC}\left(\mathbf{x}_t\right)}{\partial \mathbf{x}_t} = \frac{1}{N}\sum_{j=1}^{2}\sum_{i=1}^{N}\delta_j(l_{i,t} - \mathbf{d}_{i,t}\mathbf{x}_t)\frac{\mathbf{d}_{i,t}^T}{\sigma_j}\frac{1}{\left(1+e_{i,t}^2/\sigma_j\right)^2} = 0 \quad (32)$$

The results can be described as follows:

$$\hat{\mathbf{x}}_{t|t} = \left(\sum_{i=1}^{N}\Theta_{i,t}\left(e_{i,t}\right)\mathbf{d}_{i,t}^T\mathbf{d}_{i,t}\right)^{-1}\left(\sum_{i=1}^{N}\Theta_{i,t}\left(e_{i,t}\right)\mathbf{d}_{i,t}^T l_{i,t}\right) \quad (33)$$

$$\Theta_{i,t} = \sum_{j=1}^{2}\left(\delta_j \frac{1}{\sigma_j}\frac{1}{\left(1+e_{i,t}^2/\sigma_j\right)^2}\right) \quad (34)$$

Equation (33) can be expressed more compactly by using vector and matrix notation:

$$\hat{\mathbf{x}}_{t|t} = \left(\mathbf{D}_t^T\mathbf{\Theta}_t\mathbf{D}_t\right)^{-1}\left(\mathbf{D}_t^T\mathbf{\Theta}_t\mathbf{D}_t\right) \quad (35)$$

where $\quad \mathbf{\Theta}_t = \begin{bmatrix} \mathbf{\Theta}_{P,t} & 0 \\ 0 & \mathbf{\Theta}_{R,t} \end{bmatrix} \quad (36)$

$$\mathbf{\Theta}_{P,t} = diag\left(\Theta_{1,t}(e_{i,t}),...,\Theta_{n,t}(e_{n,t})\right) \quad (37)$$

$$\mathbf{\Theta}_{R,t} = diag\left(\Theta_{n+1,t}(e_{n+1,t}),...,\Theta_{n+m,t}(e_{n+m,t})\right) \quad (38)$$

Utilize the matrix inversion lemma, then equation (35) can be shortened to:

$$\hat{\mathbf{x}}_{t,t} = \hat{\mathbf{x}}_{t|t-1} + \tilde{\mathbf{K}}_t\left(\mathbf{y}_t - \hat{\mathbf{y}}_{t|t-1}\right) \quad (39)$$

where $\quad \tilde{\mathbf{K}}_t = \tilde{\mathbf{P}}_{t|t-1}\tilde{\mathbf{U}}_t^T\left(\tilde{\mathbf{U}}_t\tilde{\mathbf{P}}_{t|t-1}\tilde{\mathbf{U}}_t^T + \tilde{\mathbf{R}}_t\right)^{-1} \quad (40)$

$$\tilde{\mathbf{P}}_{t|t-1} = \mathbf{B}_{P,t|t-1}\mathbf{\Theta}_{P,t}^{-1}\mathbf{B}_{P,t|t-1}^T \quad (41)$$

$$\tilde{\mathbf{R}}_t = \mathbf{B}_{R,t}\mathbf{\Theta}_{R,t}^{-1}\mathbf{B}_{R,t}^T \quad (42)$$

The results obtained from Eq.(39) is the optimal estimate of $\hat{\mathbf{x}}_t$. Since $\tilde{\mathbf{K}}_t$, $\tilde{\mathbf{P}}_{t|t-1}$, and $\tilde{\mathbf{R}}_t$ are all related to $\mathbf{x}_t$, Eq.(39) is a fixed point equation about $\mathbf{x}_t$, meaning Eq.(39) can be written as $\mathbf{x}_t = \mathbf{h}(\mathbf{x}_t)$. An initial value $\hat{\mathbf{x}}_t^{(0)}$ can be given and can be solved by fixed-point iteration $\hat{\mathbf{x}}_t^{(k)} = \mathbf{h}\left(\hat{\mathbf{x}}_t^{(k-1)}\right)$.

The convergence threshold for fixed-point iteration is set as follows (where $\varepsilon$ a small normal number):

$$\left\|\hat{\mathbf{x}}_{t|t}^{(k)} - \hat{\mathbf{x}}_{t|t}^{(k-1)}\right\|/\hat{\mathbf{x}}_{t|t}^{(k-1)} \leq \varepsilon \quad (43)$$

Finally, the state error covariance matrix $\mathbf{P}_{t|t}$ is computed by:

$$\mathbf{P}_{t|t} = \left(\mathbf{I}_{n\times n} - \tilde{\mathbf{K}}_t\tilde{\mathbf{U}}_t\right)\mathbf{P}_{t|t-1}\left(\mathbf{I}_{n\times n} - \tilde{\mathbf{K}}_t\tilde{\mathbf{U}}_t\right)^T + \tilde{\mathbf{K}}_t\mathbf{R}_t\tilde{\mathbf{K}}_t^T \quad (44)$$

The pseudocode of the CKMMC-UKF algorithm is shown in Algorithm 1.

**Algorithm 1** CKMMC-UKF pseudocode

1: **Initialization**: $\hat{\mathbf{x}}_{0|0}$, $\mathbf{P}_{0|0}$, $\varepsilon$, $\alpha$, $\beta$, $\delta_1$, $\delta_2$, $\sigma_1$, $\sigma_2$
2: **for** $i = 1, 2, ...$ **do**
3:　　Calculate: $\hat{\mathbf{x}}_{t|t-1}$ and $\mathbf{P}_{t|t-1}$ by Eq.(13) and Eq.(14)
　　　　$\hat{\mathbf{y}}_{t|t-1}$ and $\tilde{\mathbf{U}}_t$ by Eq.(19) and Eq.(21)
　　　　$\mathbf{L}_t$ by $\mathbf{D}_t$ by Eq.(27) and Eq.(28)
4:　　Set $t = 1$ and $\hat{\mathbf{x}}_t^{(t=0)} = \hat{\mathbf{x}}_{t|t-1}$
5:　　Calculate: $e_{i,t} = l_{i,t} - \mathbf{d}_{i,t}\mathbf{x}_t$
　　　　$\hat{\mathbf{x}}_{t,t}$ and $\tilde{\mathbf{K}}_t$ by employing Eq.(39),Eq.(40)
6:　　If $\left\| \hat{\mathbf{x}}_t^{(t)} - \hat{\mathbf{x}}_t^{(t-1)} \right\| / \hat{\mathbf{x}}_t^{(t)} \le \varepsilon$ then $\hat{\mathbf{x}}_{t|t} = \hat{\mathbf{x}}_t^{(t)}$, go to *step 7*
　　　　else $t \to t+1$, go back to *step 5*
7:　　Update $\mathbf{P}_{t|t}$ by employing Eq.(44)
　　　　$t \to t+1$ and go back to *step 3*
8: **end for**

## IV. DERIVATION OF THE BWB-CKMMC-UKF ALGORITHM

It can be observed that using the CKMMC criteria as the optimal criterion increases the number of coefficients. However, there are coefficients such as the shape coefficients of the Cauchy kernel ($\sigma_1$, $\sigma_2$) in CKMMC and the scale coefficients that impact the selection of sigma points ($\alpha$, $\beta$) in an unscented transform that have the greatest influence on the estimation performance [21,24,26].

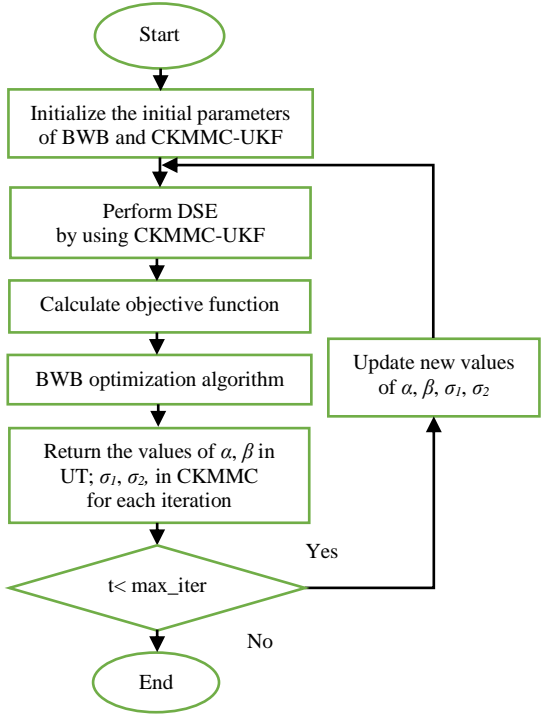

**Fig 1.** Flowchart of the BWB-CKMMC-UKF algorithm

Currently, the value of coefficients is often selected through testing and based on the experience of the algorithm designer.

Recently, using metaheuristic algorithms to optimize coefficients, and matrices or estimate missing parameters of object models has been actively discussed [28,29,31]. In this paper, an optimization algorithm is designed to deal with the above problem, and selects the best values of these coefficients that fit with the PS. The flow chart of the proposed algorithm is given in Figure 1. Beluga Whale Optimization (BWO) was introduced in 2022 by Zhong [32], describing the hunting strategy of Beluga Whale, a highly competitive algorithm. The BWO algorithm consists of three phases: exploitation, exploration, and whale fall. Two phases, exploration and exploitation, are converted through the $\mu$ coefficient.

$$\mu = \mu_0 \left[ 1 - 0.5 \left( T / T_{max} \right) \right] \quad (45)$$

where $\mu_0$: a random number in $(0,1)$; $T_{max}$: the maximum iterative number; $T$: the current iteration.

***Exploration phase:*** BWO optimized based on population, each individual (agent) will represent an optimal solution $TF(W_i)$ with the position $W_i$. In this phase, the search agents move to new positions according to this formula:

$$\begin{cases} W_{i,j}^{T+1} = W_{i,q_j}^T + \left( W_{r,q_1}^T - W_{i,q_j}^T \right)(1+c_1)\sin\left(2\pi c_2\right); j = even \\ W_{i,j}^{T+1} = W_{i,q_j}^T + \left( W_{r,q_1}^T - W_{i,q_j}^T \right)(1+c_1)\cos\left(2\pi c_2\right); j = odd \end{cases} \quad (46)$$

where $W_{i,j}^{T+1}$: the new position of $i^{th}$ Begula Whale on the $j^{th}$ dimension, $W_{i,q_j}^T$, $W_{r,q_1}^T$: the current position for $i^{th}$, $r^{th}$ Begula Whale, respectively; $q_j$ ($j=1,2,...,s$): random number selected in s-dimension. $c_1, c_2$: random numbers in $(0,1)$.

***Exploitation phase:*** the position of search agents is updated:

$$W_i^{T+1} = c_3 W_{best}^T - c_4 W_i^T + 0.1c_4 \left( 1 - \frac{T}{T_{max}} \right) \frac{\eta \times \varsigma}{|\upsilon|^{2/3}} \left( W_r^T - W_i^T \right) \quad (47)$$

where $W_{best}^T$: the best position; $\eta, \varsigma$: normally distributed random numbers; $c_3, c_4$: random number in $(0,1)$.

***Whale fall:*** When the number of whales in the population changes, the positions of search agents are updated as follows:

$$\rho = 0.1 - 0.05T / T_{max} \quad (48)$$

$$W_i^{T+1} = c_5 W_i^T - c_6 W_r^T + c_7 \left( u_b - l_b \right) \exp\left( -2\rho\phi T / T_{max} \right) \quad (49)$$

where $l_b$, $u_b$: lower and upper boundary of variables; $c_5, c_6, c_7$: random number in $(0,1)$; $\phi$: the number of whales.

Additionally, the BAT algorithm proposed by Yang in 2010 [33], is inspired by the echolocation of bats. The mathematical description of this process as follows:

$$f_i = f_{min} + (f_{max} - f_{min})\gamma \quad (50)$$

$$v_i^{t+1} = v_i^t + \left( x_i^t - x_* \right)f_i \quad (51)$$

$$x_i^{t+1} = x_i^t + v_i^t \quad (52)$$

where: $x_i$: the position of bats; $x_*$: the best position of bats; $v_i$: the velocity of bats; $f_i$; $f_{min}$; $f_{max}$: current frequency, maximum frequency, and minimum frequency of waves, respectively; $\gamma$: a random number in $[0,1]$.

It can be seen that the formula (47) involves four random values, which makes the BWO algorithm converge slowly. To enhance convergence speeds, the BAT algorithm is used in the exploitation phase of the BWO algorithm, which yields an optimal algorithm called the BWB algorithm. Based on the above analyses, the BWB-CKMMC-UKF algorithm is constructed as follows:

The first, UKF is optimized based on the CKMMC criterion, performs state estimation, and gives an expression for evaluating the average root mean square error (*ARMSE*) of the estimation. The *ARMSE* is an expression consisting of four variables: $\alpha$, $\beta$, $\sigma_1$, $\sigma_2$, which are the scale coefficients that impact the selection of sigma points in unscented transform and the shape coefficients of the Cauchy kernel.

Then, the BWB algorithm uses the ARMSE expression as the objective function (*OF*) and optimizes it under certain variable constraints. The BWB optimization iteratively finds the minimum value of the OF and the corresponding values of the four variables. Upon completion of the iterations, these coefficients are updated in the CKMMC-UKF. The coefficients obtained through BWB optimization are compatible with the power system, satisfy the Cholesky decomposition conditions, and ensure numerical stability. In this specific application, it should be noted that $TF(\mathbf{W}_i)=OF_{ARMSE}$, and $\mathbf{W}_i$ is a matrix consisting of the four coefficients to be optimized ($\mathbf{W}_i=[\ \alpha\ \beta\ \sigma_1\ \sigma_2]$).

*Remark:* The scaling coefficients alpha and beta that affect the sigma point selection in BWB-CKMMC-UKF may not be the same as those in BWB-UKF, because the two algorithms have different objective functions.

Another issue, considering Eq.(35), which involves the inverse operation, may appear singular so that additional numerical stabilization measures are required. Denoting the inverse part in Eq.(35) as $\mathbf{Z}_{inverse}=\mathbf{D}_t^T \boldsymbol{\Theta}_t \mathbf{D}_t$, then the Moore-Penrose pseudo-inverse matrix $\mathbf{Z}_{inverse}^+$, which is obtained based on singular value decomposition [34], is employed instead of the traditional inverse matrix $\mathbf{Z}_{inverse}^-$. Specifically:

$$\mathbf{Z}_{inverse}=\boldsymbol{\Lambda}\mathbf{T}\boldsymbol{\Omega}^T \qquad (53)$$

where $\mathbf{T}\in\mathbb{R}^{n\times n}$: diagonal matrix containing singular values; $\boldsymbol{\Lambda}\in\mathbb{R}^{n\times n}$, $\boldsymbol{\Omega}\in\mathbb{R}^{n\times n}$: orthogonal matrix.

The pseudo-inverse matrix $\mathbf{Z}_{inverse}^+$, in which the non-zero elements in $\mathbf{T}$ are inverted to become $\mathbf{T}^+$, is computed as:

$$\mathbf{Z}_{inverse}^+=\boldsymbol{\Omega}\mathbf{T}^+\boldsymbol{\Lambda}^T \qquad (54)$$

It can be observed that when $\mathbf{Z}_{inverse}$ is not singular then $\mathbf{Z}_{inverse}^+=\mathbf{Z}_{inverse}^-$, conversely when $\mathbf{Z}_{inverse}$ is singular then $\mathbf{Z}_{inverse}^+$ still exists. Finally, formula (35) can be rewritten as:

$$\hat{\mathbf{x}}_{t|t}=\mathbf{Z}_{inverse}^+\left(\mathbf{D}_t^T\boldsymbol{\Theta}_t\mathbf{D}_t\right) \qquad (55)$$

Combining the above modifications, the pseudocode of the BWB-CKMMC-UKF algorithm is presented in Algorithm 2.

---

**Algorithm 2** BWB-CKMMC-UKF pseudocode

---

1: **Initialization**: $\hat{\mathbf{x}}_{0|0}$, $\mathbf{P}_{0|0}$, $f_{min}$, $f_{max}$, $\delta_1$, $\delta_2$, $\varepsilon$, $v$, $x_*$, $\phi$,

   max_iterations; $W^*$ (the best solution)

2: **for** $i =1,2,...$ **do**

3:  Perform the same estimate as **Algorithm 1** and give

   the expression $OF_{ARMSE}$, in which utilize Eq.(55)

4:  *While* $l$ < max_iterations

      Calculating $\mu$, $\rho$ by Eq.(45) and (48)

    **for** each agent (*Begula Whale*)

    *If* $\mu(l) > 0.5$ (*exploration phase*)

      Updating search agents' position by Eq.(46)

    *Else if* $\mu(l) \le 0.5$ (*exploitation phase*)

       Updating search agents' position by Eq.(50-52)

    *End if* Check fitness values of search agents

    *If* $\mu(l) \le \rho$ (*whale fall*)

       Updating search agents' position by Eq.(49)

       *End if* Check fitness values of search agents

    **end for**

      Check the value $OF_{RMSE}$ and updating $W^*$, $l \rightarrow l+1$

    *End while*

5:  Updating $\alpha$, $\beta$, $\sigma_1$, $\sigma_2$; and go back to *step 3*

6: **end for**

---

## V. Experimental results

The PS parameters are shown in Table I ($k_i$: voltage amplitude; $k_v$: power injection; $k_p$: power flow, n: number of states, m: number of measurements, L: number of buses). The initial conditions such as independent Monte Carlo experiments $D = 200$; total sample time T=100; mixture coefficients $\delta_l=\delta_2=0.5$; $\varepsilon =10^{-6}$; $\mathbf{P}_{0|0}=10^{-2}$; $\mathbf{Q}_0=10^{-5}\mathbf{I}_n$ :process covariance matrices; $\mathbf{R}_0=10^{-2}\mathbf{I}_m$ : measurement covariance matrices; number of iterations in BWB algorithm: *max_iter* =500. The data and the initial state $\mathbf{x}_0$ are used the same as the website [35] to test the system and $\mathbf{E}\left[\hat{\mathbf{x}}_{0|0}\right]=\mathbf{x}_0$ ; $\mathbf{E}\left[(\mathbf{x}_0-\hat{\mathbf{x}}_{0|0})(\mathbf{x}_0-\hat{\mathbf{x}}_{0|0})^T\right]=\mathbf{P}_{0|0}$ . The simulation program is run on a Core™ i7-5600U-CPU 2.60GHz computer. To increase convincingness, BWB-CKMMC-UKF is compared with EKF, UKF, BWB-UKF, MCC-UKF [14], CKMC-UKF, and CKMMC-UKF.

The coefficient values of some algorithms are given in Table II (η: shape coefficients of the Gaussian kernel). Note that values of $\sigma_1$ and $\sigma_2$ were obtained through multiple trials, value of other coefficients is similar in references [19,21]. This paper evaluates the amplitude and phase errors of the voltage using the *ARMSE* criterion (*C*: number of buses, *D*: number of independent Monte-Carlo experiments; $\mathbf{V}_t^j$, $\hat{\mathbf{V}}_t^j$ : actual value and the estimated value of voltage amplitude, $\varphi_t^j$, $\hat{\varphi}_t^j$ : actual value and the estimated value of voltage phase respectively at the $j^{th}$ Monte-Carlo experiments ):

$$ARMSE(\mathbf{V}_t) = \sqrt{\frac{1}{CD}\sum_{j=1}^{D}\left\|\hat{\mathbf{V}}_t^j - \mathbf{V}_t^j\right\|_2^2}\Big/N \tag{56}$$

$$ARMSE(\boldsymbol{\varphi}_t) = \sqrt{\frac{1}{CD}\sum_{j=1}^{D}\left\|\hat{\boldsymbol{\varphi}}_t^j - \boldsymbol{\varphi}_t^j\right\|_2^2}\Big/N \tag{57}$$

TABLE I
PARAMETERS OF THE PS MODEL

| Parameters | 57-bus | 30-bus | 14-bus |
|---|---|---|---|
| L | 57 | 30 | 14 |
| n | 113 | 59 | 27 |
| m | 331 | 172 | 96 |
| $k_i$ | 80 | 41 | 20 |
| $k_v$ | 57 | 30 | 14 |
| $k_p$ | 57 | 30 | 14 |

TABLE II
VALUES OF COEFFICIENTS

| Algorithm \ Coefficients | $\alpha$ | $\beta$ | $p$ | $\sigma_1$ | $\sigma_2$ |
|---|---|---|---|---|---|
| UKF | 1 | 0.1 | ./. | ./. | ./. |
| MCC-UKF | 1 | 0.1 | 5 | ./. | ./. |
| CKMC-UKF | 1 | 0.1 | ./. | 1.8 | ./. |
| CKMMC-UKF | 1 | 0.1 | ./. | 1.5 | 1.5 |

### A. Performance evaluation of the BWB algorithm

For optimal algorithms (OA), their performance is usually evaluated through 23 benchmark functions (BF).

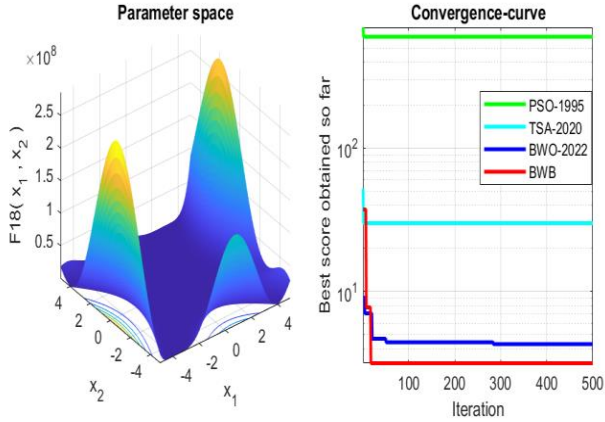

**Fig. 2.** Compare the performance of the OA

In this paper, BWB is compared to BWO, Tunicate Swarm Algorithm (TSA) [36], and Particle Swarm Optimization (PSO) [37] via 23 BF. The comparison results are shown in Figure 2. The initial condition as follows: $\phi$=30; $max\_iter$ =500; $v$=50; $f_{min}$=10; $f_{max}$=100. Survey results confirm that BWB has better optimal searching than BWO, TSA, and PSO.

### B. Performance evaluation of BWB-CKMMC-UKF algorithm

The BWB-CKMMC-UKF algorithm is tested on IEEE 57, 30, and 14-bus systems when influenced by complex conditions.

**Scenario 1:** *Laplacian noise and random outliers*

In this scenario, the noise impact on PS as follows: $\mathbf{q}_t \approx L(0,10^{-5})$ (where: $L(0,10^{-5})$ the Laplace distribution with scale $10^{-5}$ and mean 0) and $\mathbf{r}_t \approx 0.8G(0,10^{-2}) + 0.2G(0,0.5)$. It should be noted that the Gaussian distribution with a small probability and a large variance such as $G(0,0.5)$ and $G(0,10^{-2})$ can be considered outliers (impulsive noises) [13]. At the same time, set measurement covariance matrices $\mathbf{R}_t = 10^{-2}\mathbf{I}_m$, and process covariance matrices $\mathbf{Q}_t = 10^{-5}\mathbf{I}_n$. The measurement noise has random outliers $0.2G(0,0.5)$. Specifically, Fig.3 illustrates the voltage amplitude (V-A) estimation error on IEEE 30-bus, Fig.4 illustrates the voltage phase (V-P) estimation error on IEEE 57-bus, and Table III presents the V-P estimation error for each IEEE system in this scenario.

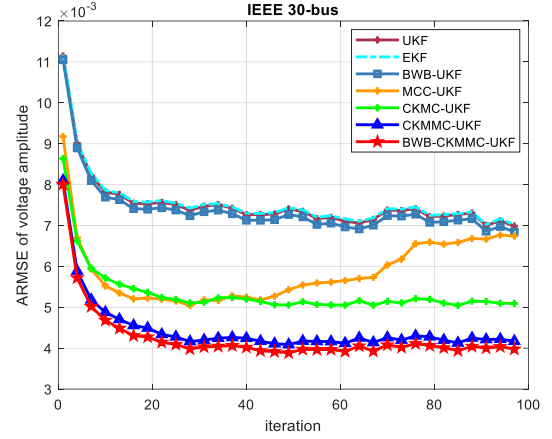

**Fig. 3.** ARMSE of V-A on IEEE 30-bus for scenario 1.

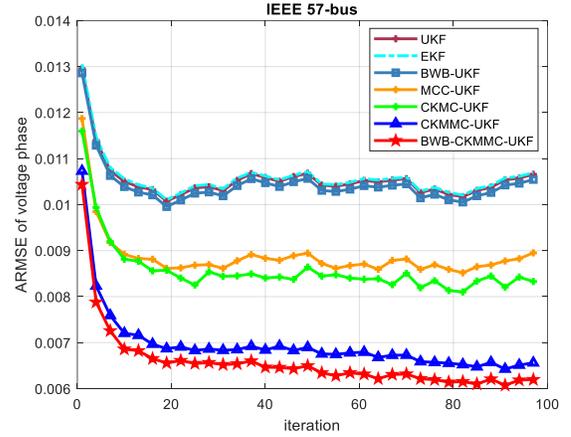

**Fig. 4.** ARMSE of V-P on IEEE 57-bus for scenario 1.

TABLE III
VOLTAGE PHASE ESTIMATION ERROR FOR SCENARIO 1

| IEEE Algorithm | Voltage phase estimation error | | |
|---|---|---|---|
| | 14-bus | 30-bus | 57-bus |
| UKF | 0.010237 | 0.010498 | 0.010528 |
| EKF | 0.010272 | 0.010546 | 0.010569 |
| BWB-UKF | 0.010118 | 0.010382 | 0.010414 |
| MCC-UKF | 0.008448 | 0.008765 | 0.008964 |
| CKMC-UKF | 0.007728 | 0.008526 | 0.009018 |
| CKMMC-UKF | 0.006311 | 0.007069 | 0.007537 |
| BWB-CKMMC-UKF | **0.005705** | **0.006687** | **0.006951** |

It is easily observed that the proposed algorithm has the smallest V-A and V-P estimation error. The estimated performance between CKMMC-UKF and CKMC-UKF has a large difference, which is due to the fusion of two Cauchy kernels which enhances the flexibility. At the same time, comparison between MCC-UKF and CKMC-UKF, the results also confirmed the stability of the Cauchy kernel compared to the Gaussian kernel under the influence of random outliers. On the IEEE 14-bus system, BWB-CKMMC-UKF gives the best results and has 9.4% better efficiency than CKMMC-UKF, 25.2% better than CKMC-UKF, and 31.8% better than MCC-UKF.

**Scenario 2**: *Bad measurement data*

In this scenario, assume that the power measurements are affected by outliers that allow the measurement to contain bad data at the 15th and 35th test times. At the 15th time, the power measurements (real power and reactive power) decreased by 10% compared to the actual time; at the 35th, it increased by 15% [3]. The noise affects the system as $\mathbf{q}_t \approx G(0,20)$ and $\mathbf{r}_t \approx 0.3G(0,25) + 0.7L(0,0.5)$. After the testing process, Fig.5 illustrates the V-A estimation error on IEEE 14-bus and Fig.6 shows the V-P estimation error in this scenario.

A comparison of the estimation results between CKMC-UKF and MCC-UKF reveals that the Cauchy kernel is more stable than the Gaussian kernel. For large-scale power systems with numerous unknown nodes, the utilization of Gaussian kernels may lead to increased estimation errors [25]. The proposed algorithm continues to provide the best estimates. Specifically, on the IEEE 57-bus system, the efficiency of the BWB-CKMMC-UKF is 31.3% higher than that of the MCC-UKF, 15.6% higher than the CKMC-UKF, and 8.2% higher than the CKMMC-UKF.

**Scenario 3**: *Load changes suddenly increase/decrease.*

In scenario 3, this paper sets out the following situations: Assuming that during the 15th and 35th test times, the power system undergoes a sudden load change, causing the voltage amplitude of node-5 to decrease by 6% and increase by 9%, respectively [24]. The noise affecting the system same as in scenario 1. After the testing process, Fig.7 illustrates the V-P estimation error on the IEEE 57-bus, and Fig.8 shows the V-A estimation error in this scenario.

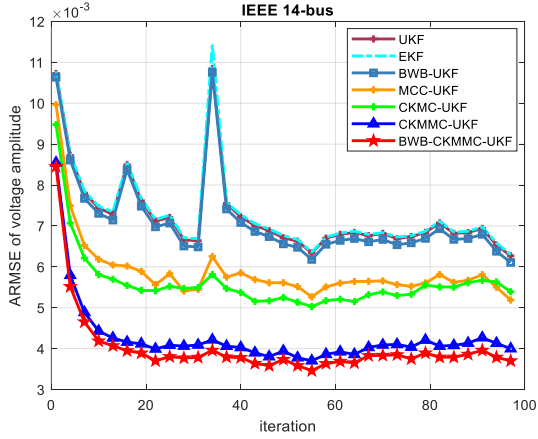

**Fig. 5.** ARMSE of V-A on IEEE 14-bus for scenario 2.

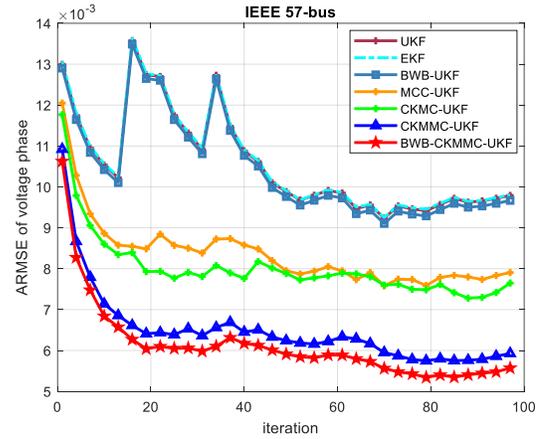

**Fig. 7.** ARMSE of V-P on IEEE 57-bus for scenario 3.

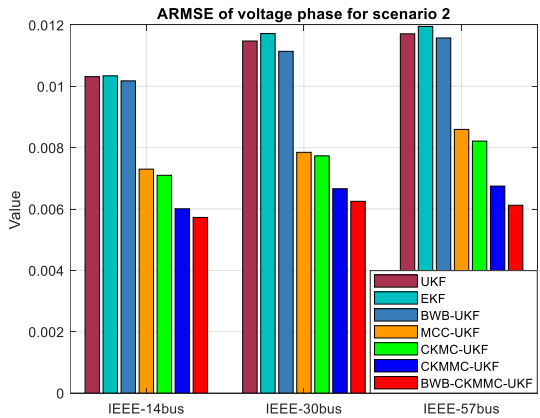

**Fig. 6.** ARMSE of V-P for scenario 2.

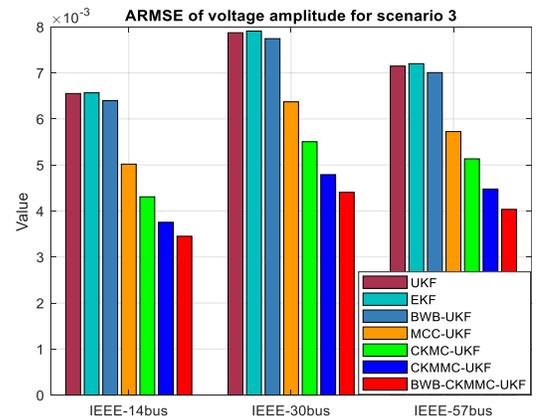

**Fig. 8.** ARMSE of V-A for scenario 3.

In a manner similar to scenario 1, the proposed algorithm consistently demonstrates outstanding performance. The EKF, UKF, and BWB-UKF filters yield estimated values with significant errors and require considerable time to achieve accurate results. In contrast, filters utilizing the correntropy criteria as optimization criteria exhibit superior performance.

The aforementioned figures demonstrate the exceptional adaptability of BWB-CKMMC-UKF in this complex scenario. Consistent with the observations in scenario 3, the advantages of the Cauchy kernel over the Gaussian kernel become increasingly pronounced as the system size increases and the impact conditions become more complex. By merging two

Cauchy kernels, the proposed algorithm effectively mitigates the challenges associated with limited measurement information. At the moment of sudden load change, BWB-CKMMC-UKF consistently provides estimates with high accuracy relative to the actual values. Specifically, in the IEEE 30-bus system, the efficiency of BWB-CKMMC-UKF is 31.3% higher than that of MCC-UKF, 20.1% higher than CKMC-UKF, and 8.1% higher than CKMMC-UKF.

### C. Computation time and computational complexity

To evaluate the algorithms proposed in this paper more comprehensively and rigorously, the computational time and complexity are provided and analyzed, in which the simulation conditions are set up the same as scenario 1. Specifically, the single-step running time of the estimation algorithms is given in Table IV, and the computational complexity is provided in Table V, respectively.

#### TABLE IV
#### COMPARE RUNNING TIMES

| IEEE Algorithm | Single-step running time (s) | | |
|---|---|---|---|
| | 14-bus | 30-bus | 57-bus |
| UKF | 0.0045 | 0.0389 | 0.1779 |
| BWB-UKF | 0.0051 | 0.0396 | 0.1808 |
| MCC-UKF | 0.0078 | 0.0445 | 0.2560 |
| CKMC-UKF | 0.0081 | 0.0447 | 0.2532 |
| CKMMC-UKF | 0.0103 | 0.0598 | 0.3091 |
| BWB-CKMMC-UKF | 0.0115 | 0.0609 | 0.3112 |

#### TABLE V
#### COMPUTATIONAL COMPLEXITY OF CKMMC-UKF

| Equation | Multiplication, addition and subtraction | Division, Cholesky decomposition, exponentiation, and matrix inversion |
|---|---|---|
| (39) | $2nm$ | 0 |
| (40) | $4mn^2 + 4m^2n - 3mn^2 + m^2$ | $O(m^3)$ |
| (41) | $4n^3 - n^2$ | $n + O(n^3)$ |
| (42) | $4m^3 - m^2$ | $m + O(m^3)$ |
| (44) | $8n^3 + 4nm^2 - 2nm + 4n^2$ | 0 |

Here, it is necessary to understand that "single-step running time" refers to the average running time for a single computational step (one iteration) of the algorithm, which includes all operations required to perform state estimation at a specific time ($t$). Each time step $t$ corresponds to collecting new measurement data from the power system and performing one state estimation loop. This loop consists of: prediction step and measurement update. It should be noted that the "single-step running time" is not the total time for the entire process. Instead, it only measures the time for one independent computational step at each time $t$, corresponding to one complete iteration. This time is averaged over many Monte Carlo experiments. For example, the CKMMC-UKF algorithm takes an average of 0.0103, 0.0598, and 0.3091 seconds to perform a full loop (from prediction to update) at each time step t on the 14, 30, and 57 bus systems, respectively.

It can be observed that the single-step running time of BWB-CKMMC-UKF is the highest, and the increase compared to CKMMC-UKF is due to the involvement of the BWB optimization algorithm. That is, the single-step computational time of BWB-CKMMC-UKF includes the time required by the BWB algorithm to determine the values of the free coefficients, as well as the time needed to perform a complete iteration. However, the trade-off between computational time and the accuracy and robustness achieved is entirely acceptable.

Furthermore, based on Table V, in which $n$ is the number of states and $m$ is the number of measurements, it can be seen that the computational complexity of CKMMC-UKF is equivalent to that of CKMC-UKF, where the computational complexity of CKMC is provided in reference [15].

### D. Investigating the effect of kernel number

In the above scenarios, the algorithms proposed based on the CKMMC criterion all employ two Cauchy kernels. To further clarify the correlation between the increase in the number of kernels and the achieved performance, the cases using two, three, and four Cauchy kernels in the CKMMC criterion are compared. Specifically, under the same noise conditions as in scenario 1, the results are illustrated in Figures 9 and 10.

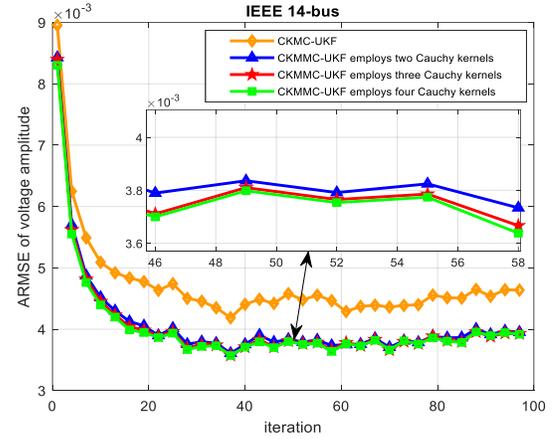

**Fig. 9.** ARMSE of V-A on IEEE 14-bus

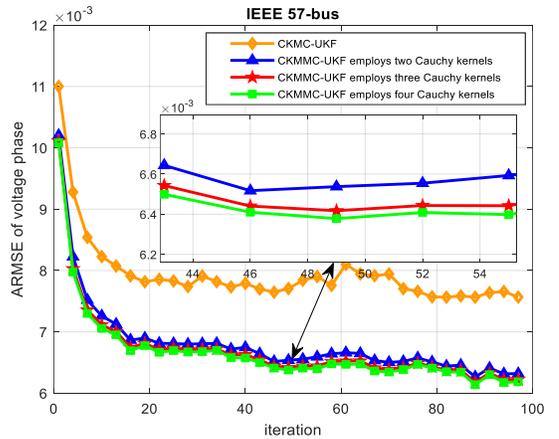

**Fig. 10.** ARMSE of V-P on IEEE 57-bus

Considering the obtained results, it can be concluded that the algorithms employing multiple kernels will achieve better performance than those employing only one kernel due to better modeling capabilities. On the other hand, theoretically, when comparing algorithms employing multiple kernels, increasing the number of kernels will bring more flexibility, thereby achieving better performance. However, the difference in performance between algorithms employing multiple kernels is not significant. In addition, increasing the number of kernels will face the challenge of increasing the shape coefficients to be selected and the computational complexity. If the shape coefficients are not well chosen, the results achieved may be lower than the algorithm employing fewer kernels but choosing better coefficients. Therefore, when considering the trade-off between the advantage gained by using multiple kernels and the difficulties that need to be solved, the choice of using the CKMMC criterion with two kernels is a reasonable choice.

## VI. CONCLUSION

In this paper, a robust BWB-CKMMC-UKF estimation algorithm has been proposed for the DSE-PS task. The proposed CKMMC optimization criterion has eliminated the limitations of the Gaussian kernel, and at the same time, well handled the problem of non-uniform distribution of data. In addition, the BWB optimization algorithm has been designed to determine the free coefficients. The optimal value of the shape coefficients of the Cauchy kernel in CKMMC criteria and the scale coefficients that impact selection sigma points in unscented transform, found through BWB optimization fit with the PS model, improved the estimation performance. It can be observed that a new and more flexible kernel function is created when two or more kernels are merged, which a single kernel cannot provide. Experimental results on IEEE 57, 30, and 14-bus systems have demonstrated the adaptability, flexibility, and excellent performance of the BWB-CKMMC-UKF algorithm compared to other variants.

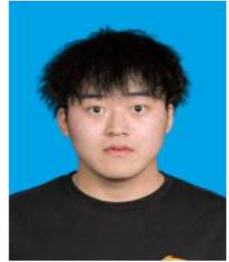

**Jinhui Hu** received the B.E. degree in electrical engineering and automation from Chang' An University, Xi'an, China, in 2022. He is currently working toward a master's degree in signal and information processing from the School of Electrical Engineering, Southwest Jiaotong University, Chengdu, China.

His current research interests include state estimation and adaptive filtering algorithms.

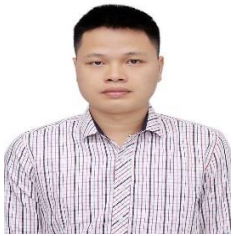

**Duc Viet Nguyen** received a master's degree in control engineering and automation from Le Quy Don University, Viet Nam, in 2019. He is working toward the Ph.D. degree in signal and information processing from the School of Electrical Engineering at Southwest Jiaotong University, Chengdu, China.

His current research interests include designing automatic control systems, state estimation, and adaptive filtering algorithms.

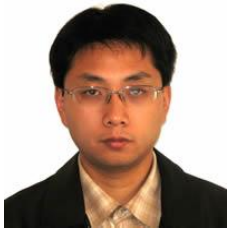

**Haiquan Zhao** (Senior Member, IEEE) received the B.S. degree in applied mathematics and the M.S. and Ph.D. degrees in signal and information processing from Southwest Jiaotong University, Chengdu, China, in 1998, 2005, and 2011, respectively.

Since 2012, he has been a Professor with the School of Electrical Engineering, Southwest Jiaotong University. From 2015 to 2016, he was a Visiting Scholar with the University of Florida, Gainesville, FL, USA. He is the author or coauthor of more than 280 international journal papers (SCI indexed) and owns 56 invention patents. His current research interests include information theoretical learning, adaptive filters, adaptive networks, active noise control, Kalman filters, machine learning, and artificial intelligence.

Dr. Zhao has won several provincial and ministerial awards and many best paper awards at international conferences or IEEE TRANSACTIONS. He has served as an Active Reviewer for several IEEE TRANSACTIONS, IET series, signal processing, and other international journals. He is currently a Handling Editor of Signal Processing, and also an Associate Editor for IEEE Transaction on Audio, Speech and Language Processing, IEEE TRANSACTIONS ON SYSTEMS, MAN AND CYBERNETICS: SYSTEM, IEEE SIGNAL PROCESSING LETTERS, IEEE SENSORS JOURNAL, and IEEE OPEN JOURNAL OF SIGNAL PROCESSING.